# Loss compensation by spasers in metamaterials

E.S. Andrianov,[1,2] D. G. Baranov,[1,2] A.A. Pukhov,[1,2] A.V. Dorofeenko,[1,2] A.P. Vinogradov,[1,2] and A.A. Lisyansky[3]

[1]Moscow Institute of Physics and Technology, 9 Institutskiy per., Dolgoprudniy 141700, Moscow Reg., Russia

[2]Institute for Theoretical and Applied Electromagnetics, 13 Izhorskaya, Moscow 125412, Russia

[3]Department of Physics, Queens College of the City University of New York, Queens, NY 11367

We show that exact loss compensation can be achieved in active metamaterials containing spasers pumped over a wide range of pumping values both below and above the spasing threshold. We demonstrate that the difference between spaser operation below and above spasing threshold vanishes, when the spaser is synchronized by an external field. As the spasing threshold loses its significance, a new pumping threshold, the threshold of loss compensation, arises. Below this threshold, which is smaller than the spasing threshold, compensation is impossible at any frequency of the external field.

## I. INTRODUCTION

Artificially created composite structures – plasmonic metamaterials – are attracting ever-growing interest thanks to their prospects for having properties not readily available in nature [1,2]. Numerous applications of metamaterials containing metallic nanoparticles (NPs) cannot be realized due to the high level of Joule losses. Gain media (atoms, molecules or quantum dots) can be incorporated into the matrix of metamaterial to compensate for losses [3-8]. The general goal of such compensation is to construct a gain metamaterial in which the electromagnetic response mimics the response of an ordinary composite without loss. In other words, such a metamaterial should be characterized by a dielectric permittivity. In particular, the wave generated by an external harmonic field in this metamaterial should have the same frequency as the external field.

The goal of loss compensation in metamaterials can be achieved with the help of spasers [9-15]. A spaser operating above the pumping threshold (the spasing regime) is an autonomic (self-oscillating) system exhibiting undamped harmonic oscillations. These oscillations are characterized by their own frequency $\omega_a$ and the amplitude [16,17]. An external harmonic field with frequency $\nu$ can synchronize the spaser forcing it to operate at frequency $\nu$ [9]. However, in this case, an active metamaterial cannot be described by a dielectric function because the synchronization is possible only if the amplitude of the external field exceeds a threshold value $E_{synch}(\Delta_E)$, which depends on the frequency detuning $\Delta_E = \nu - \omega_a$. The domain in which $E(\Delta_E) > E_{synch}(\Delta_E)$ is called the Arnold tongue [18]. In addition, for the field not much stronger than $E_{synch}(\Delta_E)$, the response of a synchronized spaser depends only weakly on the external field [10].

Generally, inside the Arnold tongue the dipole moment of the spaser has a nonzero imaginary part. This means that the spaser operates either as a gain or loss inclusion. Exact compensation occurs when the amplitude of the wave travelling in an active metamaterial is equal to a special value $E_{com}(\Delta_E)$ which depends on the frequency detuning and pump [10]. Interestingly, inside the Arnold tongue, the amplitude of the travelling wave reaches $E_{com}(\Delta_E)$ automatically. If the amplitude of the travelling wave is greater than $E_{com}(\Delta_E)$, the energy is transferred from the wave to spasers and the amplitude of the travelling wave drops. In the opposite case, the energy is transferred from the spasers to the wave and the wave amplitude grows. Thus, eventually the wave travels with the amplitude independent on the external field [14].

Below the pumping threshold, $D_{th}$, spasers are always synchronized by an external harmonic field, and therefore the system can be characterized by an effective dielectric permittivity. In this regime, spasers seem to be good candidates for loss compensation if not for the apparent energy shortage. Indeed, self-oscillations of a spaser above the pumping threshold are due to the energy delivered by pumping which cannot be smaller than Joule losses. An external field causes additional losses, whose compensation requires increased pumping. Since below the threshold, the pumping does not compensate for losses even in the absence of an external field, it seems that in order to achieve compensation, one should pump the driven spaser above threshold. These arguments qualitatively agree with the result obtained in Ref. [13] where it was shown that loss compensation occurs simultaneously with the start of spasing. On the other hand, in the absence of the interaction between plasmonic and active media [7], as well as in the linear regime [11,12], compensation the below-threshold was predicted.

In this paper, we demonstrate that the exact compensation of Joule losses can be achieved with spasers in both below and above the pumping threshold. We show that the synchronizing external field destroys self-oscillations of spasers transforming them into a nonlinear oscillator with a new pumping threshold $D_{comp}(E) < D_{th}$. When the pumping rate, $D_0$, exceeds $D_{comp}$, there is a line of exact loss compensation, $E = E_{comp}(\omega_a - \omega_E, D_0)$.

## II. EQUATIONS OF MOTIONS FOR A FREE SPASER AND A SPASER DRIVEN BY EXTERNAL OPTICAL WAVE

We consider a simplest model of spaser as a two level system (TLS) of size $r_{TLS}$ placed near a plasmonic spherical NP of size $r_{NP}$ [16]. The energy from the pumped TLS is non-radiatively transferred to the NP exciting surface plasmons (SPs). At the frequency of the SP resonance, the dynamics of the NP dipole moment is governed by the oscillator equation:

$$\ddot{\mathbf{d}}_{NP} + \omega_{SP}^2 \mathbf{d}_{NP} = 0. \tag{1}$$

The quantization of this oscillator can be performed in an ordinary way by introducing the Bose operators $\hat{\bar{a}}^\dagger(t)$ and $\hat{\bar{a}}(t)$ for the creation and annihilation of the dipole SP [19,20]:

$$\hat{\mathbf{d}}_{NP} = \boldsymbol{\mu}_{NP}\left(\hat{\bar{a}} + \hat{\bar{a}}^\dagger\right).$$

The corresponding Hamiltonian is:

$$\hat{H}_{SP} = \frac{\hbar\omega_{SP}}{2}\left(\hat{\bar{a}}^\dagger\hat{\bar{a}} + \hat{\bar{a}}\hat{\bar{a}}^\dagger\right). \tag{2}$$

To determine the value $\boldsymbol{\mu}_{NP}$, we should equate the energy of a single plasmon to the energy of the quant:

$$\frac{\hbar\omega_{SP}}{2} = \frac{1}{8\pi}\int \left.\frac{\partial \operatorname{Re}(\omega\varepsilon(\omega))}{\partial \omega}\right|_{\omega_{SP}} |\mathbf{E}_{NP}|^2 dV$$
$$= \frac{1}{8\pi}\int \left(\operatorname{Re}\varepsilon(\omega) + \omega\frac{\partial \operatorname{Re}\varepsilon(\omega)}{\partial \omega}\right)\bigg|_{\omega_{SP}} |\mathbf{E}_{NP}|^2 dV, \tag{3}$$

where permittivity $\varepsilon = \varepsilon_{NP}$ inside the NP and $\varepsilon = 1$ in the surrounding volume, and $\mathbf{E}_{NP}$ is the electric field of the NP. In the absence of the external field, for the NP near field we have $(8\pi)^{-1}\int \operatorname{Re}\varepsilon\big|_{\omega_{SP}}|\mathbf{E}_{NP}|^2 dV = 0$, where $\varepsilon = \varepsilon_{NP}$ inside the NP and $\varepsilon = 1$ in the surrounding volume [21]. The near field is not equal to zero at the resonance frequency only. However, in this case, the field vanishes at the infinity and $\int \text{Re}\,\varepsilon\big|_{\omega_{SP}} |\mathbf{E}_{NP}|^2 dV = \int \text{Re}\,\varepsilon\big|_{\omega_{SP}} \nabla\varphi^* \cdot \nabla\varphi \, dV$
$= -\int \varphi^* \nabla \cdot \left( \text{Re}\,\varepsilon\big|_{\omega_{SP}} \nabla\varphi \right) dV$. The latter integral is equal to zero both inside and outside the NP. Furthermore, $\partial \text{Re}\,\varepsilon_{NP}(\omega)/\partial\omega = 0$ outside of the particle. This modifies Eq. (3) as

$$\frac{\hbar\omega_{SP}}{2} = \frac{1}{8\pi} \int_{\substack{volume \\ of\ NP}} \omega \frac{\partial \text{Re}\,\varepsilon_{NP}(\omega)}{\partial\omega}\bigg|_{\omega_{SP}} |\mathbf{E}_{NP}|^2 dV, \quad (3a)$$

For a spherical NP with the radius $r_{NP}$, the electric field of the SP with a unitary dipole moment, $\boldsymbol{\mu}_1$, is equal to $\mathbf{E}_{NP} = -\boldsymbol{\mu}_1 r^{-3} + 3(\boldsymbol{\mu}_1 \cdot \mathbf{r})\mathbf{r} r^{-5}$ and $\mathbf{E}_{NP} = -\boldsymbol{\mu}_1 r_{NP}^{-3}$ outside and inside of the NP, respectively. Thus, we obtain

$$\frac{\hbar\omega_{SP}}{2} = \frac{|\boldsymbol{\mu}_{NP}|^2}{6 r_{NP}^3} \omega_{SP} \frac{\partial \text{Re}\,\varepsilon}{\partial\omega}\bigg|_{\omega_{pl}}, \quad (4)$$

which gives

$$\boldsymbol{\mu}_{NP} = \sqrt{3\hbar r_{NP}^3 \Big/ \left( \frac{\partial \text{Re}\,\varepsilon_{NP}}{\partial\omega} \right)} \frac{\boldsymbol{\mu}_1}{|\boldsymbol{\mu}_1|}. \quad (5)$$

The quantum description of a TLS is done via the transition operator $\hat{\tilde{\sigma}} = |g\rangle\langle e|$ between ground $|g\rangle$ and excited $|e\rangle$ states of the TLS, so that the operator for the dipole moment of the TLS is represented as

$$\hat{\boldsymbol{\mu}}_{TLS} = \boldsymbol{\mu}_{TLS} \left( \hat{\tilde{\sigma}}(t) + \hat{\tilde{\sigma}}^\dagger(t) \right), \quad (7)$$

where $\boldsymbol{\mu}_{TLS} = \langle e|e\mathbf{r}|g\rangle$ is the TLS dipole moment matrix element. The Hamiltonian of the two-level TLS can be written as

$$\hat{H}_{TLS} = \hbar\omega_{TLS} \hat{\tilde{\sigma}}^\dagger \hat{\tilde{\sigma}}, \quad (8)$$

where $\omega_{TLS}$ is the transition frequency of the TLS.

The commutation relations for operators $\hat{\tilde{a}}(t)$ and $\hat{\tilde{\sigma}}(t)$ are standard: $\left[ \hat{\tilde{a}}(t), \hat{\tilde{a}}^+(t) \right] = 1$ and $\left[ \hat{\sigma}^\dagger, \hat{\sigma} \right] = \hat{D}$, where the operator $\hat{D} = \left[ \hat{\sigma}^\dagger, \hat{\sigma} \right] = \hat{n}_e - \hat{n}_g$ describes the population inversion of the ground $\hat{n}_g = |g\rangle\langle g|$ and excited states $\hat{n}_e = |e\rangle\langle e|$, $\hat{n}_g + \hat{n}_e = \hat{1}$, of the TLS.

We describe the dynamics of the free spaser by the model Hamiltonian[8,16,22]

$$\hat{H} = \hat{H}_{SP} + \hat{H}_{TLS} + \hat{V} + \hat{\Gamma}, \quad (9)$$

where the operator $\hat{V} = -\hat{\mathbf{d}}_{NP} \hat{\mathbf{E}}_{TLS}$ is responsible for the dipole-dipole interaction between the TLS and the NP. Taking into account that $\hat{\mathbf{E}}_{TLS} = -\hat{\boldsymbol{\mu}}_{TLS} r^{-3} + 3(\hat{\boldsymbol{\mu}}_{TLS} \cdot \mathbf{r})\mathbf{r} r^{-5}$, we obtain

$$\hat{V} = \hbar\Omega_R \left(\hat{\tilde{a}}^\dagger + \hat{\tilde{a}}\right)\left(\hat{\tilde{\sigma}}^\dagger + \hat{\tilde{\sigma}}\right). \tag{10}$$

where $\Omega_R = \left[\boldsymbol{\mu}_{NP} \cdot \boldsymbol{\mu}_{TLS} - 3(\boldsymbol{\mu}_{TLS} \cdot \mathbf{e}_r)(\boldsymbol{\mu}_{NP} \cdot \mathbf{e}_r)\right]/\hbar r_0^3$ is the Rabi frequency, $r_0$ is the distance between the TLS and the NP, and $\mathbf{e}_r = \mathbf{r}/r$ is the unitary vector. The last term in the Hamiltonian (9) is responsible for relaxation and pumping processes.

Assuming that $\omega_{TLS}$ is close to the frequency of the plasmonic resonance, $\omega_{SP} \approx \omega_{TLS}$, we can use the approximation of the rotating wave [23] by looking for the solutions in the form $\hat{\tilde{a}}(t) \equiv \hat{a}(t)\exp(-i\omega_a t)$ and $\hat{\tilde{\sigma}}(t) \equiv \hat{\sigma}(t)\exp(-i\omega_a t)$, where $\hat{a}(t)$, $\hat{\sigma}(t)$ are slow varying in time operators and $\omega_a$ is the autonomous frequency of the spaser which we seek. Disregarding fast-oscillating terms proportional to $\exp(\pm 2i\omega_a t)$, the interaction operator $\hat{V}$ may be written the form of the Jaynes–Cummings Hamiltonian [20]:

$$\hat{V} = \hbar\Omega_R(\hat{a}^\dagger \hat{\sigma} + \hat{\sigma}^\dagger \hat{a}), \tag{11}$$

Using Hamiltonian (9) and the commutation relations for operators $\hat{a}(t)$ and $\hat{\sigma}(t)$ we obtain the Heisenberg equations of motion for the operators $\hat{a}(t)$, $\hat{\sigma}(t)$, and $\hat{D}(t)$ [22]

$$\dot{\hat{D}} = 2i\Omega_R(\hat{a}^\dagger \hat{\sigma} - \hat{\sigma}^\dagger \hat{a}) + 2i\Omega_2(\hat{\sigma} - \hat{\sigma}^\dagger) - \left(\hat{D} - \hat{D}_0\right)\tau_D^{-1}, \tag{12}$$

$$\dot{\hat{\sigma}} = \left(i\delta - \tau_\sigma^{-1}\right)\hat{\sigma} + i\Omega_R \hat{a}\hat{D} + i\Omega_2 \hat{D}, \tag{13}$$

$$\dot{\hat{a}} = \left(i\Delta - \tau_a^{-1}\right)\hat{a} - i\Omega_R \hat{\sigma} - i\Omega_1. \tag{14}$$

where $\delta = \omega_a - \omega_{TLS}$ and $\Delta = \omega_a - \omega_{SP}$ are frequency detunings. To take into account relaxation processes (the $\Gamma$-term in Eq. (9)), we phenomenologically added terms proportional to $\tau_D^{-1}, \tau_\sigma^{-1}$, and $\tau_a^{-1}$ in Eqs. (12)-(14). The term $D_0$ describes pumping [20,23] and corresponds to the population inversion in the TLS in the absence of the NP. From now on, we neglect quantum fluctuations and correlations and consider $\hat{D}(t)$, $\hat{\sigma}(t)$, and $\hat{a}(t)$ as c-numbers. In this case, the Hermitian conjugation turns into the complex conjugation [3,17,22,24]. Note that the quantity $D(t)$ that describes the difference in populations of excited and ground states of the TLS is a real quantity because the corresponding operator is Hermitian.

The system of Eqs. (12)-(14) has stationary solutions, which depend on the pumping level $D_0$. For $D_0$ smaller than the threshold value

$$D_{th} = \left[1 + (\omega_{SP} - \omega_a)^2 \tau_a^2\right] / \left(\Omega_R^2 \tau_a \tau_\sigma\right) \tag{15}$$

there is only the trivial solution $a = \sigma = 0$, $D = D_0$. For $D_0 > D_{th}$ the second stationary solution arises. In this case, the trivial solution corresponding to the absence of SPs is unstable, while the stable solution corresponds to laser generation of SPs (spasing) with the frequency [22]

$$\omega_a = (\omega_{SP}\tau_a + \omega_{TLS}\tau_\sigma)/(\tau_a + \tau_\sigma). \tag{16}$$

The Hamiltonian of a spaser driven by an external field of optical wave, which is assumed to be classical, $E_{OW}(t) = E\cos\nu t$, may be written in the form (see for details Refs. [9,10,25]):

$$\hat{H}_{eff} = \hat{H} + \hbar\Omega_1\left(\hat{\tilde{a}}^\dagger + \hat{\tilde{a}}\right)\left(e^{i\nu t} + e^{-i\nu t}\right) + \hbar\Omega_2\left(\hat{\tilde{\sigma}}^\dagger + \hat{\tilde{\sigma}}\right)\left(e^{i\nu t} + e^{-i\nu t}\right), \tag{17}$$

where $H$ is given by Eq. (9), $\Omega_1 = -\mu_{NP}E/\hbar$ and $\Omega_2 = -\mu_{TLS}E/\hbar$ are the coupling constants of the external field interaction with the NP and the TLS, respectively.

As above, the equations of motion for slow amplitudes $\hat{a}$, $\hat{\sigma}$, and $\hat{D}$ can be obtained as:

$$\dot{\hat{D}} = 2i\Omega_R(\hat{a}^\dagger\hat{\sigma} - \hat{\sigma}^\dagger\hat{a}) + 2i\Omega_2(\hat{\sigma} - \hat{\sigma}^\dagger) - \tau_D^{-1}\left(\hat{D} - \hat{D}_0\right), \tag{18}$$

$$\dot{\hat{\sigma}} = (i\delta_E - \tau_\sigma^{-1})\hat{\sigma} + i\Omega_R\hat{a}\hat{D} + i\Omega_2\hat{D}, \tag{19}$$

$$\dot{\hat{a}} = (i\Delta_E - \tau_a^{-1})\hat{a} - i\Omega_R\hat{\sigma} - i\Omega_1. \tag{20}$$

where $\Delta_E = \nu - \omega_{SP}$ and $\delta_E = \nu - \omega_{TLS}$. In the next section, using Eqs. (18)-(20) we demonstrate that spasers below the pumping threshold can be used for Joule loss compensation.

### III. LOSS COMPENSATION

Since the energy flows through a spaser, it can be considered as an open system. The flow starts at pumping, which causes the population inversion of the TLS, then the nonradiative transition of the TLS excites SPs at the NP, and the Joule losses of SPs at the NP finalize the flow. In the absence of an external field, the energy of pumping is consumed by SP excitations. An increase of the SP amplitude is limited by Joule losses at the NP. The self-oscillating state of a spaser (spasing) occurs at exact compensation of losses with pumping [13,26]. If losses exceed the energy supplied by pumping, the stationary amplitudes of oscillations are equal to zero. The maximum value of pumping, below which there is no spasing, is referred to as the threshold pumping.

The external field performs work on the dipole moments of the TLS and the NP. Thus, subjecting a spaser to an external field leads to additional channels of energy flow; namely, the energy flows from the field to the TLS and to the NP. These energy flows may close up via the interaction of the TLS with the NP and interfere with each other and with the primary energy flow from pumping to the TLS and then to Joule losses at the NP. Rather complicated dynamics of these flows results in non-zero oscillations of the below-threshold spaser, as shown in Fig. 1(a).

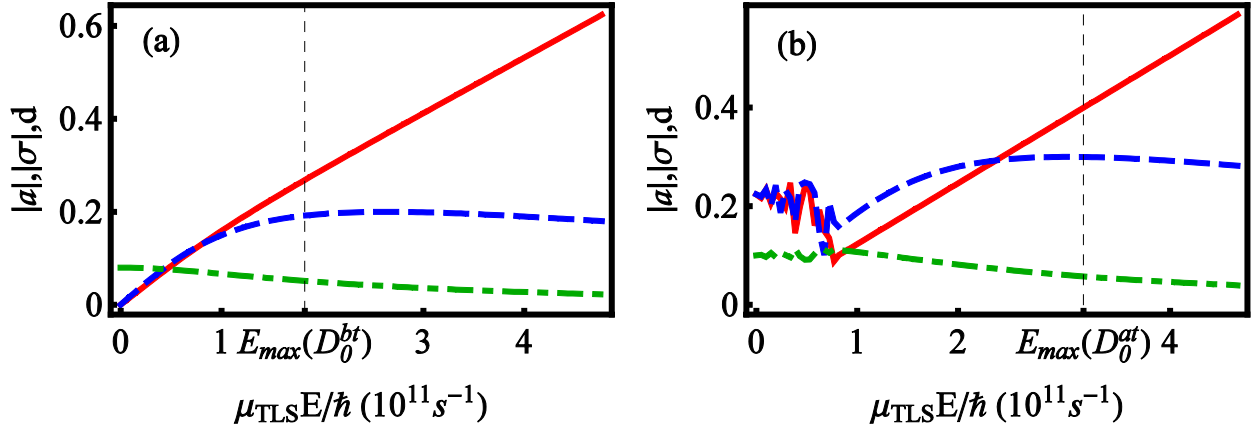

FIG. 1. The dependencies of spaser's parameters (in arbitrary units) on the amplitude of the external field (a) below and (b) above the pumping threshold $D_{th} = 0.1$. The NP dipole moment, the TLS dipole moment, and the inversion are shown by solid, dashed, and dot-dashed lines, respectively. The values of pumping for Figs. (a) and (b) are $D_0^{bt} = 0.07$ and $D_0^{at} = 0.12$, respectively. The irregular behavior of NP and TLS dipole moments at small fields corresponds to spaser stochastic oscillations outside the Arnold tongue [9]. Note, that the ratio of $E_{\max}(D_0)$ below and above threshold ($\mu_{TLS} E_{\max}(D_0^{bt})/\hbar = 0.8 \cdot 10^{12} s^{-1}$ and $\mu_{TLS} E_{\max}(D_0^{at})/\hbar = 1.1 \cdot 10^{12} s^{-1}$, respectively) is of the order of the ratio of corresponding pumping values.

For exact loss compensation, the work performed by the field on the spaser should be equal to zero. The time averaged work performed by the external field $\mathbf{E}$ on the spaser is[27] $\sim (\boldsymbol{\mu}''_{TLS} + \boldsymbol{\mu}''_{NP}) \cdot \mathbf{E}^*$. Since $\boldsymbol{\mu} \parallel \mathbf{E}$, for non-zero dipole moments, this expression turns to zero when $(\boldsymbol{\mu}''_{TLS} + \boldsymbol{\mu}''_{NP}) = 0$ or the phase difference between the field and spaser dipole oscillations is equal to $\pi$. If the phase difference is greater than $\pi$, the wave is amplified (negative work). If the phase is smaller than $\pi$, the wave attenuates (positive work). In other words, for exact loss compensation the sum of the imaginary parts of the dipole moments of the TLS and the NP should be equal to zero. It has been shown [10], that above the spasing threshold, the imaginary part of the dipole moment of a synchronized spaser is equal to zero on the compensation curve $E = E_{comp}(\Delta_E, D_0)$ (see Fig. 2). For the fields below this curve, the energy pumped into the system exceeds losses; for the fields above the compensation curve, the system becomes lossy. Note, that exact compensation can occur only for the fields smaller than $E_{max}(D_0) = \max\{E = E_{comp}(\Delta_E, D_0)\}$.

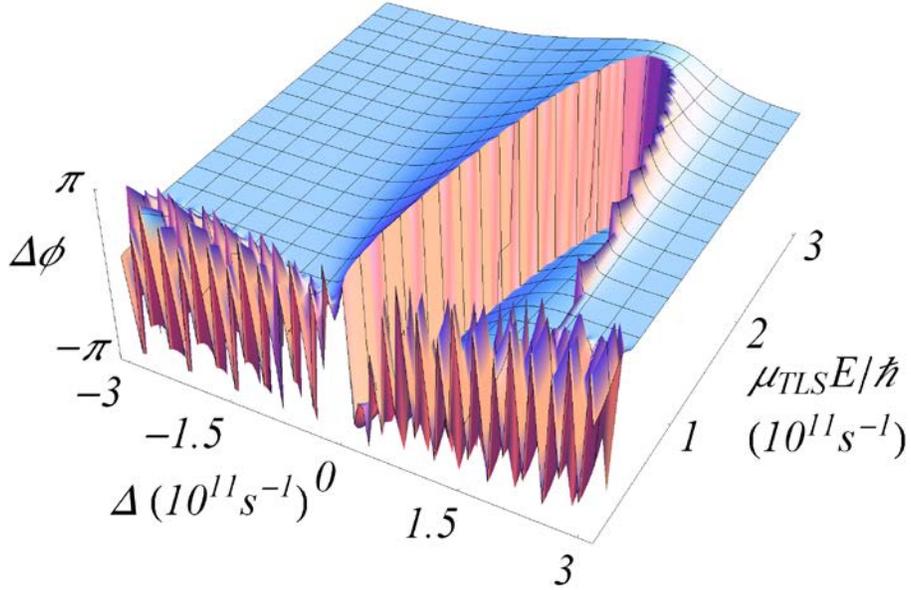

FIG. 2. The dependence of $\phi = \tan^{-1}(\operatorname{Im} d_{spaser} / \operatorname{Re} d_{spaser})$ on the amplitude of the external field $E$ and the detuning $\Delta_E$ for the above-threshold pumping. The smooth part of the surface corresponds to the Arnold tongue where the spaser is synchronized by the external field while the speckle structure at low field corresponds to the spaser's stochastic behavior. At the discontinuity line, on which $\phi = \pi$, loss is exactly compensated. In this and subsequent figures, except for the value of pumping, $D_0 = 0.12$, the calculations are made for the following values of parameters: $\tau_a = 10^{-14} s$, $\tau_\sigma = 10^{-11} s$, $\tau_D = 10^{-13} s$, and $\Omega_R = 10^{13} s^{-1}$.

The possibility of loss compensation below threshold can be illustrated in a simple limiting case of a vanishing interaction between the TLS and the NP. The pumping threshold depends on the coupling constant $\Omega_R$ (see Eq. (15)). For $\Omega_R \to 0$ the threshold tends to infinity. Such a spaser cannot spase. In such a TLS-NP couple, the TLS and the NP are independent and play opposite roles when they interact with the external wave. While the TLS amplifies the wave's field, losses in the NP weaken it. In the limit $\Omega_R = 0$, Eqs. (18)-(20) are reduced to:

$$\dot{\hat{D}} = 2i\Omega_2(\hat{\sigma} - \hat{\sigma}^\dagger) - (\hat{D} - \hat{D}_0)\tau_D^{-1}, \tag{21}$$

$$\dot{\hat{\sigma}} = (i\delta_E - \tau_\sigma^{-1})\hat{\sigma} + i\Omega_2\hat{D}, \tag{22}$$

$$\dot{\hat{a}} = (i\Delta_E - \tau_a^{-1})\hat{a} - i\Omega_1. \tag{23}$$

The stationary solution of these equations is:

$$a_{st} = \frac{(-\Delta_E + i\tau_a^{-1})\mu_{NP}/\hbar}{\Delta_E^2 + \tau_a^{-2}} E, \tag{24}$$

$$\sigma_{st} = \frac{D_0(\delta_E - i\tau_D^{-1})\mu_{TLS}/\hbar}{\tau_\sigma^{-2} + \delta_E^2 + 4(\mu_{TLS}E/\hbar)_2^2 \tau_D \tau_\sigma^{-2}} E, \tag{25}$$

$$D_{st} = \frac{D_0(\delta_E^2 + \tau_\sigma^{-2})}{\tau_\sigma^{-2} + \delta_E^2 + 4(\mu_{TLS}E/\hbar)_2^2 \tau_D \tau_\sigma^{-1}}. \tag{26}$$

The sum of imaginary parts of the TLS and NP dipoles moments, which are proportional to $\text{Im}\,a$ and $\text{Im}\,\sigma$, respectively, vanishes if

$$D_0 = \left(\frac{\mu_{NP}}{\mu_{TLS}}\right)^2 \frac{\tau_\sigma}{\tau_a} \frac{\tau_\sigma^{-2} + \delta_E^2 + 4(\mu_{TLS}E/\hbar)_2^2 \tau_D \tau_\sigma^{-1}}{\Delta_E^2 + \tau_a^{-2}}, \tag{27}$$

$$-\Delta_E + \delta_E \tau_\sigma/\tau_a < 0. \tag{28}$$

In this case, the contradiction with the energy shortage is resolved if we notice that the pumping energy should not compensate for Joule losses in the NP caused by the TLS field. The energy of pumping is transferred to the field by the TLS. At the same time the NP absorbs the field energy. If Eqs. (27) and (28) are satisfied, then the total energy transfer to the system is zero. Note that since the pumping threshold in this toy model is infinity, loss compensation occurs below the pumping threshold.

In a sense, the spaser with $\Omega_R = 0$ is similar to the system suggested in Ref. [7] in which gain and plasmonic media are confined to different layers of one-dimensional photonic crystal. Thus, there is no direct interaction between the media and, as a consequence, no spasing. The lasing starts when the energy delivered by pumping exceeds loss at plasmonic layers.

Having $\Omega_R = 0$ is sufficient but not necessary for loss compensation. As it is shown in Fig. 3(b), in the case $\Omega_R \neq 0$ below the spasing threshold, there are frequencies for which the spaser's dipole moment is zero and the energy is not transferred to or from the system. There are also frequencies for which the imaginary part of the NP dipole moment is negative, so that the NP releases energy to the wave. The reason of such unexpected behavior is interference of energy fluxes similar to the case of the Fano resonance [28].

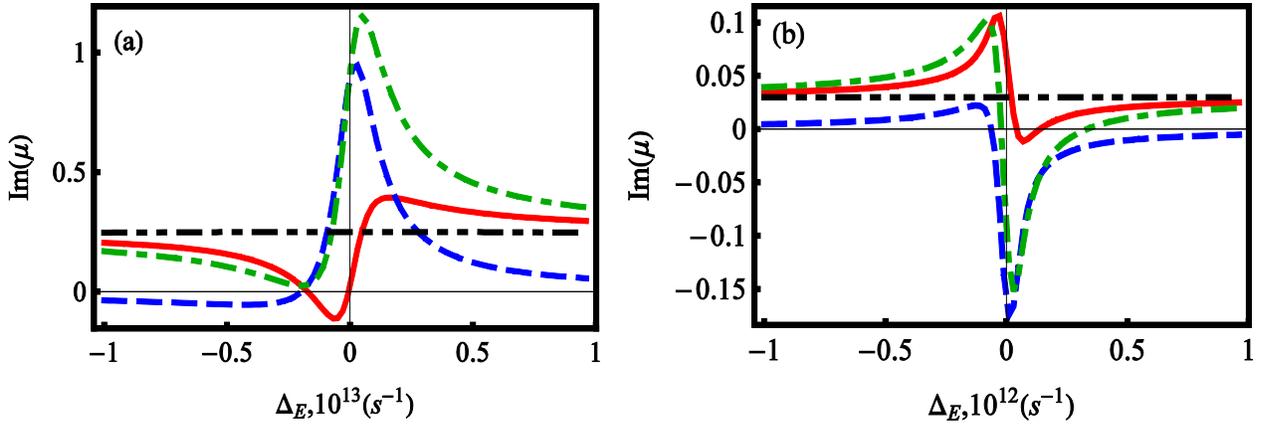

FIG. 3. The dependencies of imaginary parts of dipole moments of the whole spaser, which has $D_{th} = 0.1$, on the frequency detuning (dot-dashed lines) in the external field for the level of pumping of (a) $D_0 = -0.9$ and (b) $D_0 = 0.08$. Solid and dashed lines show imaginary parts of dipole moments of the NP and the TLS, respectively. The imaginary part of the dipole moment of the NP not interacting with the TLS in the external field is shown by the double dot-dashed line. This dependency is very slow and looks like a horizontal line at the scale of the figure.

The analogy with the Fano resonance clearly manifests itself for low pumping ($D_0 \sim -1$) when there is still no loss compensation (Fig. 3a). In this case, the spaser response to the external field is practically linear and we can consider the system as two coupled resonators. The first resonance with a low $Q$-factor is the SP resonance at the NP, the other is the high-$Q$ resonant transition of the TLS. The response of such a system on the external force has the shape of the Fano curve[29]

$$f = \frac{(\omega + q)}{\omega^2 + 1}, \qquad (29)$$

where $q$ describes asymmetry of the line.

In the case of the spaser with a low level of pumping, $D_0 \sim -1$, at $\tau_a \ll \tau_\sigma$, the frequency dependence of the imaginary part of the total dipole moment of the spaser can be obtained as

$$\operatorname{Im}\mu_{tot} \approx \frac{\left(\Delta_E + \Omega_R |D_0| \mu_{TLS} / \mu_{NP}\right)^2 + |D_0| / \tau_a \tau_\sigma}{\left(\Delta_E^2 \tau_a^2 - \left(1/\tau_a \tau_\sigma + \Omega_R^2 |D_0|\right)\right)^2 + \Delta_E^2}, \qquad (30)$$

Though the explicit dependence on $\Delta_E$ differs from Eq. (29), it qualitatively reproduces the Fano curve. The asymmetry factor $q$ is equal to $\Omega_R |D_0| \mu_{TLS} / \mu_{NP}$, which strongly depends on the interaction of the NP and the TLS with the external field.

For low pumping, $\operatorname{Im}\mu_{tot} > 0$ (see Fig. 3a), so that the system is lossy for any frequency. At the same time, the minimum of losses is significantly lower than that for an isolated NP. When pumping increases, the resonant line still resembles the Fano resonance line but the minimum value of $\operatorname{Im}\mu_{tot}$ becomes negative (Fig. 3b), so that for a range of frequencies the spaser releases energy. The compensation curve, $E = E_{comp}(\Delta, D_0)$, also exists in this case (see Fig. 4). As in the case of the above-threshold spaser, this curve lies below some value of the electric field $E_{max}(D_0)$. The main difference between spasers in the above and below threshold pumping regimes is that in the latter case, the Arnold tongue occupies the whole half-plane, i.e., the spaser is always synchronized. Below threshold, the synchronized spaser is not a self-oscillating system but it is rather a non-linear oscillator. Thus, again we arrive at the contradiction of the shortage of the pumping energy [13,30,31].

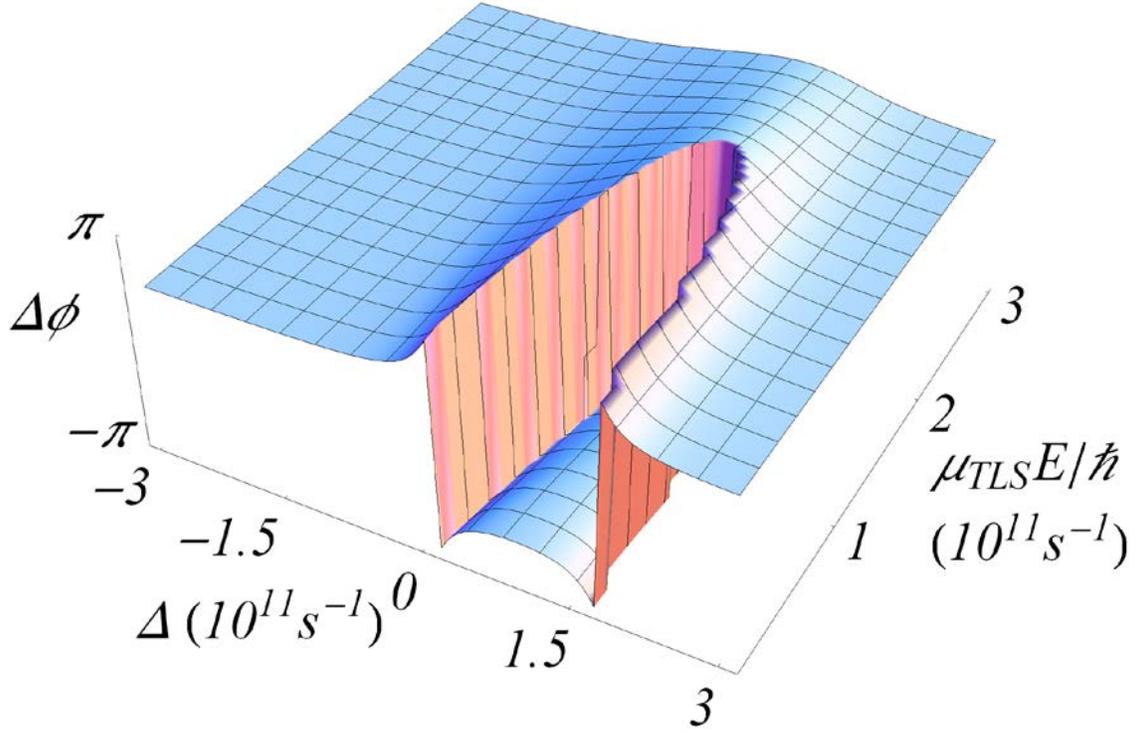

FIG. 4. The dependence of $\phi = \tan^{-1}(\operatorname{Im} d_{spaser} / \operatorname{Re} d_{spaser})$ on the amplitude of the external field $E$ and the frequency detuning $\Delta_E$ in the below-threshold pumping, $D_0 = 0.08$. The Arnold tongue occupies the whole half-plane, so that the spaser is always synchronized by the external field. At the discontinuity line, on which $\phi = \pi$, loss is exactly compensated.

This contradiction is resolved, if one notices that for the fields in which loss compensation exists ($E < E_{\max}(D_0)$) the absolute values of the dipole moments of the above-threshold spaser is greater than that for the below-threshold spaser (see Fig. 1). As a result, below pumping threshold, losses of the TLS field in the NP are smaller than that above threshold. Thus, the energy of pumping is sufficient to compensate for these losses as well as for additional Joule losses due to the external field. If $E > E_{\max}(D_0)$, the dipole moments of a spaser above and below threshold are nearly the same. As a consequence, for these fields pumping greater than the threshold is needed for loss compensation. Therefore, $E_{\max}(D_0)$ is a critical amplitude of the external field, for which the dipole moments of the below-threshold spaser become comparable with those of the *free* above-threshold spaser. If the external field exceeds $E_{\max}(D_0)$, losses in the NP of the field generated by the TLS exceed the energy supplied by pumping and the energy of pumping below the threshold becomes insufficient for loss compensation.

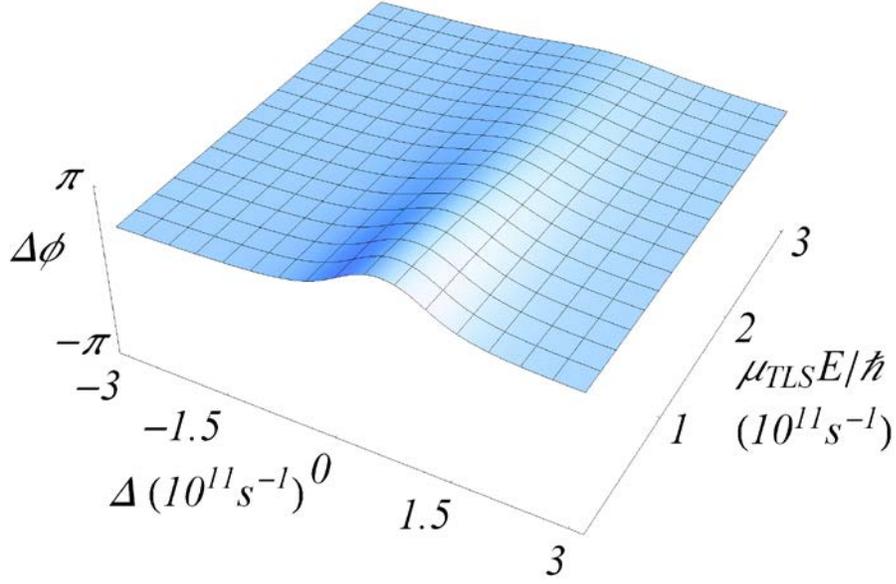

FIG. 5. The dependence of $\phi = \tan^{-1}(\operatorname{Im} d_{spaser} / \operatorname{Re} d_{spaser})$ on the amplitude of the external field $E$ and the detuning $\Delta_E$ in the below-threshold pumping, $D_0 = 0.05$. The pumping is insufficient for loss compensation, so that the compensation line doesn't exist.

As pumping decreases, the compensation curve monotonically shrinks toward the line $E = 0$ disappearing at some level of pumping (Fig. 5). The dependence $E_{\max}(D_0)$, shown in Fig. 6, is characterized by the new pumping threshold, $D_{comp}$, below which no compensation is possible. As shown in Fig. 7, $D_{comp}$ is never greater than the spasing threshold.

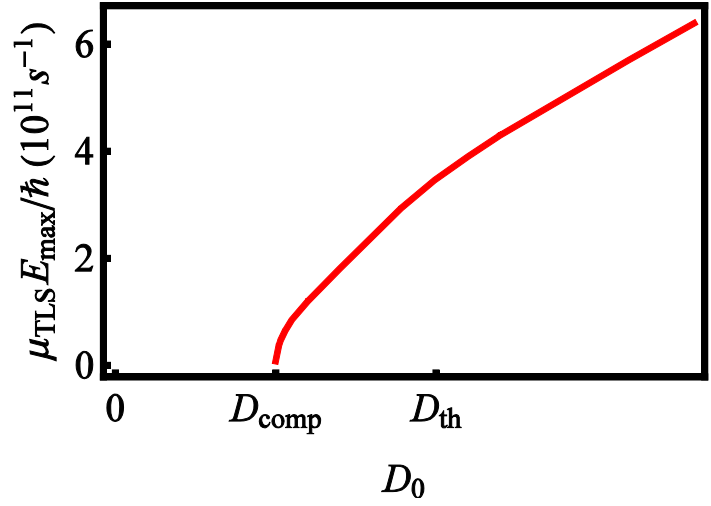

FIG. 6. The maximum value of the external field at which exact compensation takes place as a function of pumping.

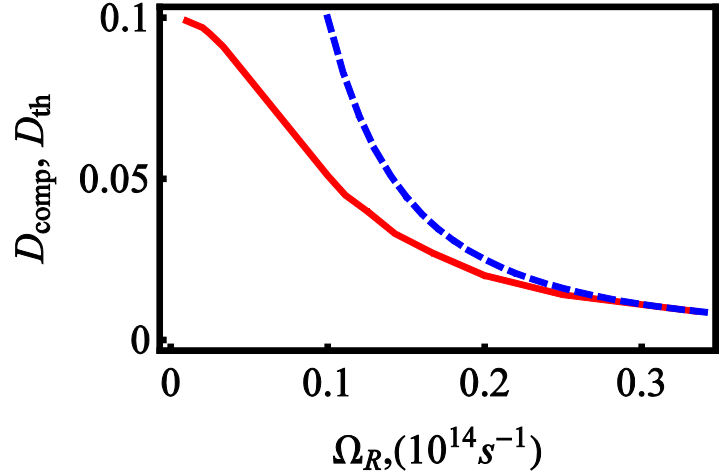

FIG. 7. The dependencies of $D_{comp}$ (solid line) and $D_{th}$ (dashed line) on $\Omega_R$.

Above and below threshold, the compensation line originates from the points $E = 0, \Delta = 0$ and $E = 0, \Delta > 0$, respectively (see Fig. 8). The pumping $D_0 = D_{th}$ is the smallest pumping at which compensation at zero frequency detuning is possible. This is the case considered in Refs. [13,26]. For $D_{comp} \leq D_0 < D_{th}$, compensation can only be achieved for $\Delta > 0$.

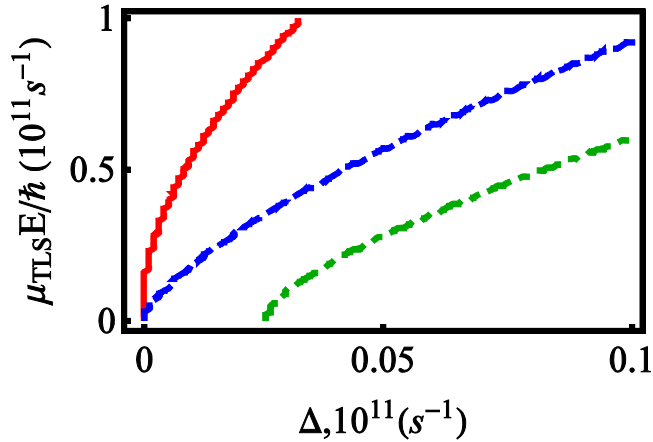

FIG. 8. Part of the compensation curves for small frequency detuning. Solid, dashed, and dot-dashed line correspond to the pumping above ($D_0 = 0.12$), at ($D_0 = D_{th} = 0.1$), and below ($D_0 = 0.08$) the threshold, respectively.

## IV. CONCLUSION AND DISCUSSION

Synchronizing a spaser to an external field leads to the destruction of the spaser as a self-oscillating system and to its transformation into an active nonlinear driven oscillator. Indeed, after synchronization, the qualitative differences between below- and above-threshold spasers disappear; the remaining differences are merely quantitative (i.e., the ratio of dipole moments is of the order of the ratio of pumping values).

The threshold pumping $D_{th}$ of a free spaser loses its significance. In particular, the exact compensation of losses by a spaser is realized over a wide range of pumping values above and below $D_{th}$. A new pumping threshold, the threshold of compensation $D_{comp}$, arises. Below this threshold compensation is impossible at any frequency of the external field.

The values of pumping, at which the exact loss compensation is achieved, and at which the threshold pumping coincide with $D_{comp}$ is in the absence of the external field and zero frequency detuning only. In this case the compensation threshold coincides with the spasing threshold in agreement with results of Ref. [13].

Below threshold, loss compensation by a spaser is only possible if the frequency of the external field is greater than the transition frequency for the TLS. In this connection, it is interesting to consider the results of Ref. [12], in which it is shown numerically that below-threshold compensation is achieved at frequencies below the transition frequency. Nevertheless, if we take into account the Lorentz shift of the resonant frequency $\Delta\omega \sim \sqrt{4\pi/3 N e^2/m}$ (see, e.g., Ref. [32]) appearing due to the difference between local and average fields (in Ref. [12] the concentration of active molecules is $N \sim 6 \cdot 10^{18} cm^{-3}$), we obtain $\Delta\omega \sim 3 \cdot 10^{13} s^{-1}$. This is in good agreement with our conclusion that below-threshold compensation is possible for positive detuning only. Thus, there is no contradiction between the results of Refs. [12] and [13].

In conclusion, when pumping is below the spasing threshold, spasers may be used for compensation for Joule losses over a range of frequencies once the necessary pumping exceeds a new compensation threshold $D_{comp}$, which is smaller than the spasing threshold.

## ACKNOWLEDGEMENTS

This work was supported by RFBR Grants No. 12-02-01093 and by a PSC-CUNY grant.